# An Instrument for Physical Vapor Deposition onto Cryo-EM Samples for Microsecond Time-Resolved Cryo-EM

Wyatt A. Curtis, Constantin R. Krüger,[†] Axel P. Tracol Gavard, Jakub Wenz,[†] Marcel Drabbels,[†] and Ulrich J. Lorenz[*]

**Affiliation:**

Ecole Polytechnique Fédérale de Lausanne (EPFL), Laboratory of Molecular Nanodynamics, CH-1015 Lausanne, Switzerland

[†] Current affiliation: Ecole Polytechnique Fédérale de Lausanne (EPFL), Laboratory for Ultrafast X-ray Sciences, CH-1015 Lausanne, Switzerland

[*] To whom correspondence should be addressed. Email: ulrich.lorenz@epfl.ch

**Abstract**

Laser flash melting and revitrification experiments have recently improved the time resolution of cryo-electron microscopy (cryo-EM) to the microsecond timescale, making it fast enough to observe many of the protein motions that are associated with function. The technique has also opened up a new dimension for cryo-EM sample preparation, making it possible to deposit compounds onto a cryo-EM sample while it is frozen, so that upon flash melting, the embedded particles experience an altered environment. For example, we have recently shown that depositing ultrathin silicon dioxide membranes onto a cryo-EM sample causes particles to detach from the interface upon flash melting, removing preferred particle orientation. Here, we describe an apparatus for physical vapor deposition of compounds onto cryo-EM samples, detailing its design and operation. As a demonstration, we determine that the minimum thickness of silicon dioxide sealing membranes in a laser flash melting experiment is just over two monolayers. Moreover, we show that it is possible to deposit reagents onto the cryo-EM and induce mixing during flash melting, which points towards a new strategy for initiating protein dynamics in time resolved experiments. We propose that the design of our instrument can form the basis for an integrated platform for microsecond time-resolved cryo-EM experiments.

Physical vapor deposition has long been used in the preparation of electron microscopy samples. In Scanning Electron Microscopy, thin metal layers are sometimes evaporated onto the sample at an angle, which creates a pseudo-three-dimensional effect.[1,2] Carbon or metal coatings are deposited onto non-conducting samples to prevent charging under illumination with the electron beam.[3] For example, a thin layer of platinum is frequently applied to cryo-electron tomography samples, which reduces artifacts during focused ion beam milling and improves image quality.[4] It is however rare for compounds to be purposely deposited onto samples for single-particle cryo-EM, which risks reducing image contrast and obscuring high-resolution features of the embedded proteins.

Microsecond time-resolved cryo-EM experiments based on laser flash melting have recently added a new dimension to cryo-EM that makes the deposition of compounds onto frozen samples a useful tool.[5-7] A cryo-EM sample is briefly melted with a microsecond laser pulse, and protein dynamics are initiated, before the laser is switched off, and revitrification traps the proteins in their transient configurations. Besides enabling time-resolved obervations, these experiments have also opened up new possibilities for improving the properties of conventional cryo-EM samples. While the buildup of water vapor on a cryo-EM sample is usually a nuisance that reduces contrast, we have recently shown that deposition of a layer of amorphous ice prior to flash melting provides an approach for overcoming preferred particle orientation,[8] an issue that plagues many cryo-EM projects.[9–12] The additional layer of ice separates particles from the air-water interface, to which they have adsorbed, so that they are able to rotate freely upon flash melting. Revitrification then traps the particles in an improved angular distribution. More recently, we have also shown that the deposition of amorphous ice followed by laser flash melting may enable a new approach for preparing cryo-EM samples with soft-landing mass spectrometry. While the soft-landed particles have been dehydrated in the vacuum of the mass spectrometer, deposition of amorphous ice and subsequent flash melting returns a fraction of the particles to a configuration close to their native structure even within the brief time window provided by a 30 μs laser pulse.[13,14]

An even wider range of applications becomes available by depositing non-volatile compounds onto cryo-EM samples with physical vapor deposition. We have recently shown that cryo-EM samples can be enclosed in an ultrathin liquid cell geometry by vapor depositing silicon dioxide membranes onto the sample, as illustrated in Fig. 1a,b.[15] These hydrophilic membranes replace the hydrophobic air-water interface, so that upon laser melting, particles detach from the interface and randomize their orientations (Fig. 1c). Upon revitrification, a near-isotropic angular distribution is obtained (Fig. 1d). Moreover, the sealing membranes prevent evaporation and stabilize the thin liquid film. This makes it possible to flash melt the sample multiple times and extend the time window in microsecond time-resolved cryo-EM experiments from tens to hundreds of microseconds. This has allowed us to study how the conformational landscape of the 50S ribosomal subunit evolves over several hundred microseconds in response to a temperature jump induced by the melting laser. These experiments also suggest that vapor deposition of reagents onto the sample should provide a new method for initiating protein dynamics.[15] Upon flash melting, the deposited compound, for example a small molecule or ligand, should rapidly mix with the sample, since ions or small molecules diffuse across the thickness of a typical cryo-EM within microseconds, and induce conformational dynamics. This will make a much broader range of dynamics accessible, beyond those that can be triggered with photocaged compounds.[16,17] Here, we describe a vacuum instrument that enables the physical vapor deposition of compounds onto cryo-EM samples and thus renders such experiments possible.

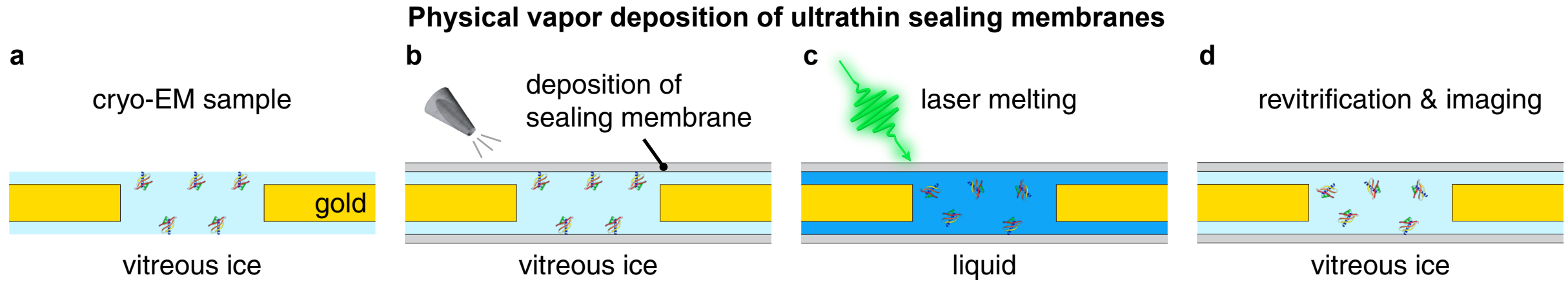

**Figure 1 | Physical vapor deposition onto cryo-EM samples — ultrathin sealing membranes for microsecond time-resolved cryo-EM experiments using laser flash melting. a,b** Thin layers of silicon dioxide are deposited onto both sides of a cryo-EM sample. **c** When the sample is flash melted with a microsecond laser pulse, these sealing membranes prevent evaporation and stabilize the thin liquid film. **d** Once the laser is switched off, the sample rapidly cools and revitrifies, with the particle orientations scrambled.

Figure 2a illustrates the vapor deposition instrument, with a cross section shown in Fig. 2b. Cryo-EM samples are inserted into the vacuum chamber with a single tilt cryo specimen holder, and components are deposited either by leaking a volatile compound into the chamber through a gas dosing valve (right) or through physical vapor deposition using an electron beam evaporator (left). The operation can be observed through a viewport on the top. The vacuum chamber is constructed from a six-way stainless steel cube with conflat flanges (CU100, Hositrad) and is evacuated to a pressure of $3 \cdot 10^{-8}$ mbar with a turbomolecular pump (685 L/s, HiPace 700, Pfeiffer) that is placed underneath the chamber and that is backed by an oil-free dry scroll pump ($1 \cdot 10^{-3}$ mbar backing pressure, nXDS20i, Edwards). The chamber pressure is monitored with a full-range pressure gauge (PKR 251, Pfeiffer) located on the flange opposite the load lock (Fig. 2b).

**Illustration of the vapor deposition instrument**

**Figure 2 | Illustration of the vapor deposition instrument. a** Overview. The cryo-EM sample is mounted in a side-entry cryo holder and inserted into the chamber through a load-lock repurposed from a JEOL 2010 transmission electron microscope. Thin sealing layers are deposited onto the cryo-EM sample with an electron beam evaporator. The chamber is also equipped with a gas dosing valve that can be used to introduce volatile compounds. **b** Cross-section of the vapor deposition chamber. The cryo-EM sample is placed at a distance of 100 mm from the electron beam evaporator and can be rotated around the axis of the cryo holder, making it possible to deposit on both sides of the sample. A quartz crystal microbalance, located opposited the evaporator, is used to calibrate the deposition rate.

Cryo-EM samples are loaded into a cryo-transfer holder (~100 K sample temperature, Elsa, Gatan) and inserted into the vacuum chamber through a load-lock that we repurposed from a retired JEOL 2010 transmission electron microscope. The load-lock is integrated into a motorized rotation stage that was previously part of the goniometer of the microscope. This allows us to rotate the sample holder around its axis and position the sample such that either side faces the electron beam evaporator during deposition. The load-lock is pumped to a pressure of $1 \cdot 10^{-5}$ mbar by a turbomolecular pump (60 L/s, TMU 071P, Pfeiffer), which is backed by another oil-free dry scroll pump ($1 \cdot 10^{-3}$ mbar backing pressure, XDS 5, Edwards). The following procedure is adopted to minimize the deposition of water vapor onto the sample during transfer. The cryo-transfer holder is initially inserted into the vacuum chamber with its cryo-shield covering the sample. For physical vapor deposition experiments, the cryo-shield is then immediately opened. Otherwise, radiative heating by the evaporator causes ice that has accumulated on the inside of the cryo-shield to sublimate and redeposit onto the colder cryo-EM sample. The cryo-shield is only closed once the deposition is complete. As the holder is removed from the vacuum chamber, the airlock is vented with dry nitrogen.

Volatile compounds can be deposited onto the cryo-EM sample by leaking them into the chamber through a gas dosing valve (All-metal gas dosing valve, 420VED016, Pfeiffer), located on the right flange in Fig. 2b. The deposition rate can be controlled by adjusting the leak rate and monitoring the pressure in the vacuum chamber. Once the chamber pressure has stabilized, the cryo-shield of the sample holder is then opened for the desired deposition time. Note that the deposition rate is lower than predicted by the Hertz-Knudsen equation,[18] depending on how much the geometry of the cryo holder shields the sample (for the Elsa holder, the deposition rate is lower by about a factor of 60, where the actual sample thickness was determined with electron energy loss spectroscopy[19–21]). In our instrument, deposition of volatile compounds that are leaked into the vacuum chamber through the gas dosing valve occurs on both sides of the sample simultaneously.

We deposit low vapor pressure compounds onto the sample through physical vapor deposition, using the electron beam evaporator (e-flux Mini E-Beam Evaporator, TecTra), which is mounted on the left flange of the chamber (Fig. 2b). A tantalum crucible with a graphite liner is filled with a fine powder of the compound and placed into the evaporator, where it is heated with an electron beam. A tungsten coil (0.2 mm diameter wire) surrounding the crucible is resistively heated with a current of about 7 A, causing it to emit electrons. The crucible is biased to a potential of +1.8 kV, so that the emitted electrons are accelerated toward the crucible to heat it by electron impact. The material inside the crucible is vaporized, and the vapor escapes through a 1 mm diameter aperture, forming an effusive beam that is deposited onto the cryo-EM sample. A shutter in front of the evaporator initially blocks the beam when the cryo-EM sample is loaded. The sample is then rotated to face the effusive beam, and the shutter is briefly opened to deposit the desired thickness. For some applications, such as for enclosing the sample between ultrathin sealing membranes, the sample is then rotated 180°, and the process is repeated to deposit the compound on the other sample side as well.

The physical vapor deposition rate (typically 10–30 Å/min) is actively stabilized during the deposition process and is calibrated with a quartz crystal microbalance (QCM). A small fraction of the effusive beam escaping from the electron beam evaporator is ionized by the electron beam, and the ions are collected by a stainless steel electrode located behind the effusive beam aperture (-24 V bias). The resulting ion current is proportional to the evaporant flux and provides a feedback signal that is used to stabilize the deposition rate by adjusting the heating current that is passed through the tungsten coil. We calibrate the deposition rate with a QCM (QM20, McVac Manufacturing) that faces the electron beam evaporator and is placed about 1.5 cm behind the sample location, about 10° off axis (Fig. 2b). Figure 3a illustrates a typical experiment, in which we measure the deposition rate with the QCM as a function of the ion current. When we open the shutter in front of the electron beam evaporator to allow the effusive beam to pass, the QCM crystal is radiatively heated, which causes the measured deposition rate to briefly become negative, before it stabilizes after about 20 s. Upon closing the shutter, a positive spike is observed. Figure 3b shows that for typical deposition rates, the rate measured with the QCM increases approximately linearly with the ion current. This allows us to rapidly set a desired deposition rate simply by adjusting the ion current. We find that due to the radiative heating of the crystal, the measured deposition rate is about 40 % higher than the actual rate. Note that we do not observe that the cryo-EM samples devitrify during the deposition process, indicating that they do not heat up significantly.

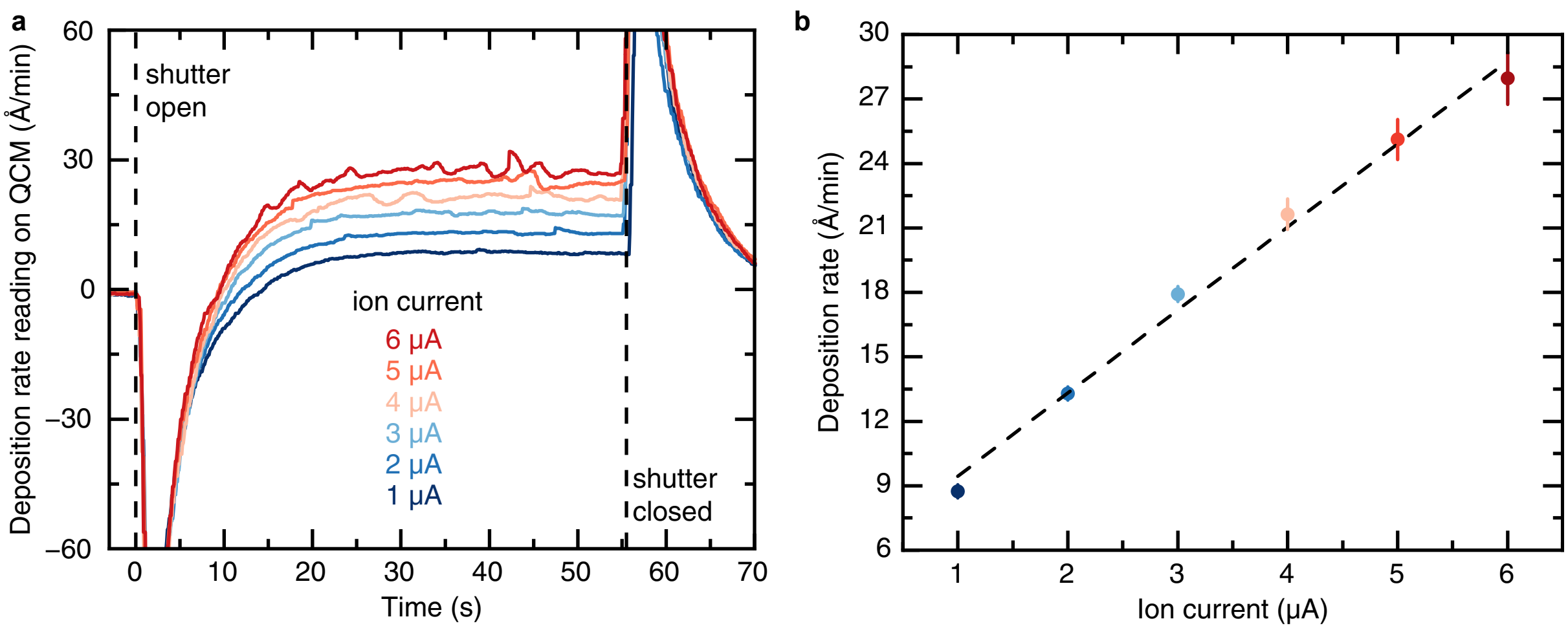

**Figure 3 | Determination of the deposition rate. a** Deposition rate measured by the QCM as a function of time for different ion current readings in the electron beam evaporator. Opening and closing the shutter of the evaporator alters the heat load on the QCM and briefly perturbs the reading, which stabilizes after about 20 s. **b** Average deposition rate as a function of the ion current measured in the electron beam evaporator (dots, error bars represent the standard deviation in a time window from 30.6 s to 52.3 s, 218 data points). The deposition rate increases linearly with the ion current (linear fit indicated with dashed line).

We demonstrate the operation of our vapor deposition instrument by using it to seal cryo-EM samples between ultrathin silicon dioxide membranes. While we have previously shown that ~1.4 nm thick membranes (about 4 monolayers) are thin enough to allow for high-resolution reconstructions,[15] it is tempting to try and reduce the membrane thickness even further in order to improve the available contrast, which should be particularly beneficial for imaging small proteins. Using the elastic mean free paths, we can estimate that the sealing membranes reduce the contrast by about as much as an ice layer of twice their thickness.[22,23] Here, we determine the minimum thickness of a silicon dioxide membrane required in a laser flash melting experiment. The crucible of the electron beam evaporator is filled with finely ground silicon dioxide (Kurt J. Lesker), and oxygen is leaked into the chamber during deposition (partial pressure of $10^{-5}$ mbar) in order to prevent the deposition of oxygen deficient species and improve the homogeneity of the sealing layers. Silicon dioxide membranes of ~0.75 nm thickness (about 2 monolayers) are deposited onto both sides of a plunge frozen test cryo-EM sample on a holey gold specimen support (1.2 µm holes, 1.3 µm apart on 300 mesh gold, Quantifoil). The sample is then melted and revitrified with a 30 µs laser pulse as previously described[15] (532 nm wavelength, about 80 mW laser power, 22 µm diameter spot size in the sample plane, with the laser beam aimed at the center of a grid square). The micrograph in Fig. 4a reveals that the membranes have withstood the flash melting process and are able to seal the sample. In contrast, reducing the membrane thickness even further to ~0.5 nm (less than 2 monolayers) results in a sample that shows strong contrast variations after laser flash melting (Fig. 4b). This indicates that the sample has partially evaporated through pores in the membrane, which causes the observed thickness variations.

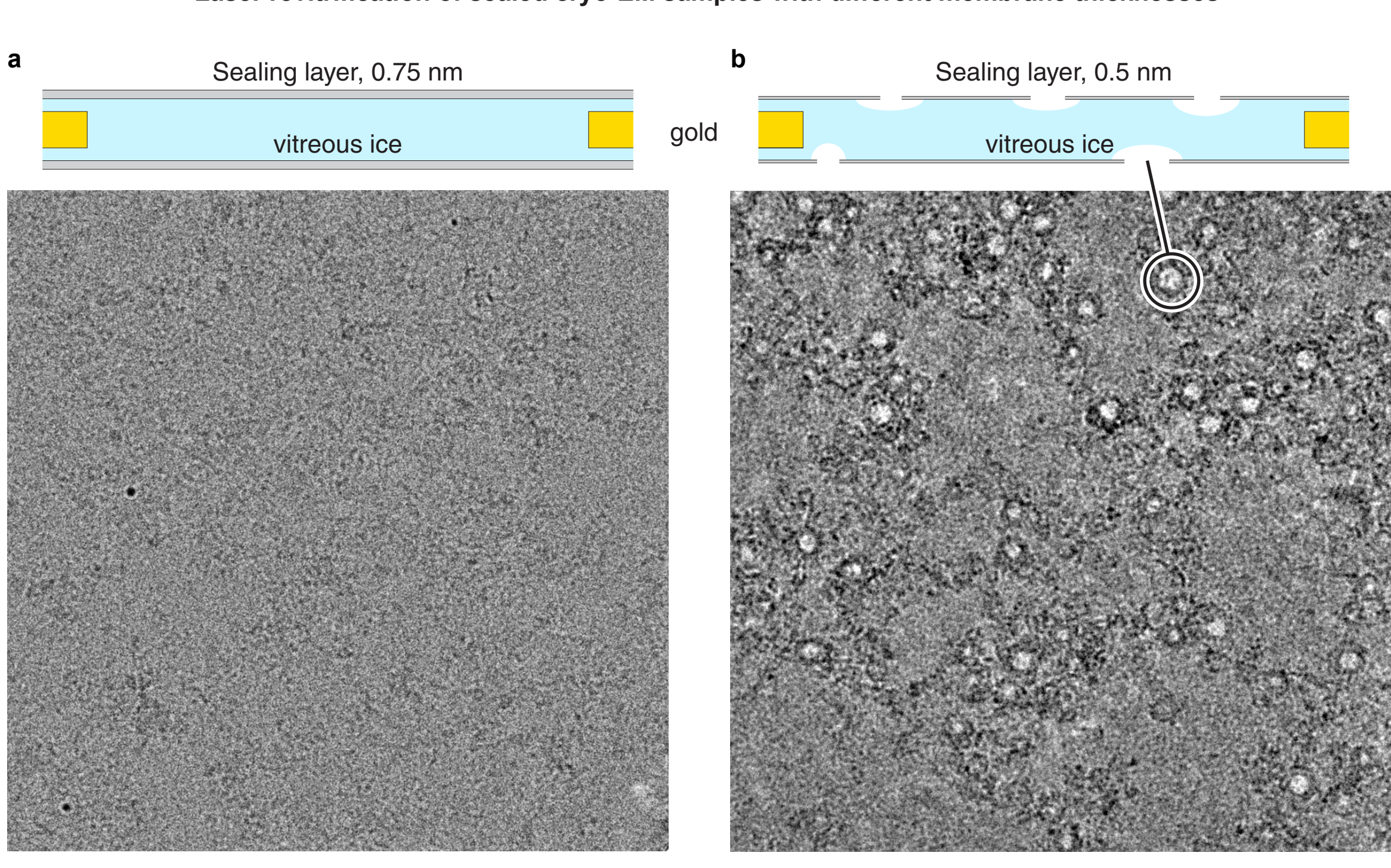

**Figure 4 | Laser revitrification of cryo-EM samples sealed between silicon dioxide layers of different thickness. a** Micrograph of a cryo-EM sample sealed between ~0.75 nm thick silicon dioxide membranes (about 2 monolayers) after revitrification with a 30 µs laser pulse. **b** When the membrane thickness is reduced to ~0.5 nm (less than 2 monolayers), evaporation of the sample through pores in the sealing membranes occurs, as indicated by the areas of lighter contrast.

Finally, we demonstrate that it is possible to perform microsecond mixing experiments by vapor depositing reagents onto a cryo-EM sample and inducing mixing with laser flash melting. We prepare a test cryo-EM sample containing the calcium sensing dye Fura Red (potassium salt, 3 mM concentration, AAT Bioquest) on a holey gold specimen support (1.2 µm holes, 1.3 µm apart on 300 mesh gold, Quantifoil), and deposit a 0.5 nm thick layer of anhydrous calcium chloride (Sigma-Aldrich) onto one side of the sample. In the presence of calcium ions, the fluorescence quantum yield of Fura Red decreases.[24] We therefore expect that in the area that is flash melted by the laser beam, the flurorescence intensity should decrease if mixing has occurred. Laser melting and revitrification experiments are performed with an optical microscope (DM6000 CFS, Leica) that we have modified for flash melting experiments as described previously.[7] The sample is placed in a cryo stage (77 K temperature, CMS 196, Linkam), where it is illuminated with a blue LED (410 nm), and micrographs are recorded in reflection with a CCD camera (acA1920-40gm, Basler). Fura Red absorbs blue light (maximum at 435 nm) and emits red light (maximum at 639 nm),[25] which allows us to record fluorescence micrographs by inserting a 600 nm long-pass filter into the microscope.

Figure 5a shows fluorscence micrographs of a typical sample area before and after flash melting with a 50 µs laser pulse (532 nm wavelength, about 50 mW laser power, 25 µm FWHM beam diameter in the sample plane, with the laser beam aimed at the center of a grid square). The fluorescence intensity noticeably decreases in the revitrified area, which is particulary evident in the difference micrograph in the inset, indicating that mixing and binding of the calcium ions to the dye have occurred. For reference, Fig. 5b shows micrographs of the same area recorded in reflection without long-pass filter. A control experiment in Fig. 5c,d shows that the contrast change in the fluorescence micrograph does not occur without deposition of calcium chloride prior to revitrification.

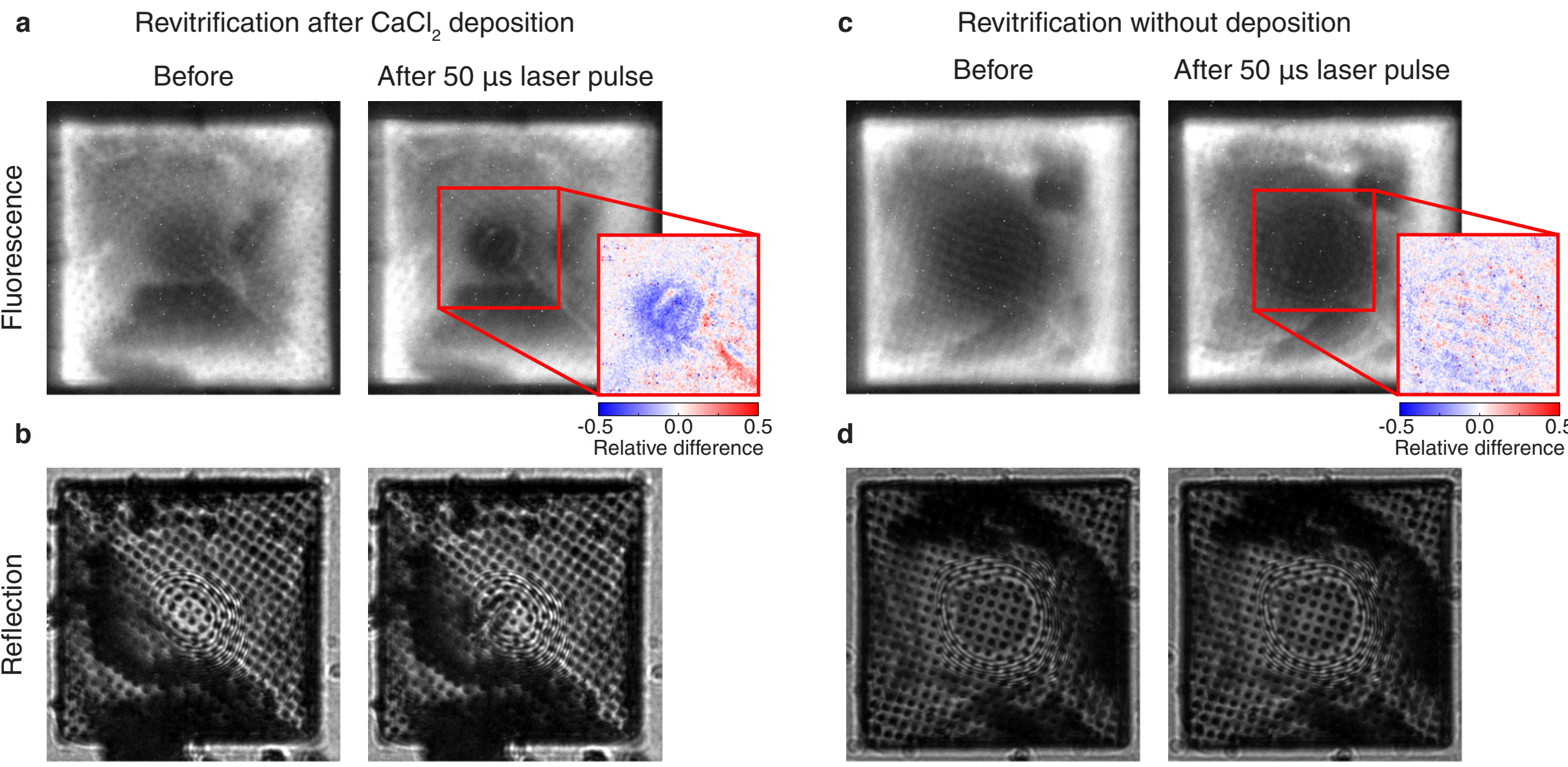

**Figure 5 | Microsecond mixing experiments with vapor deposited compounds. a** Fluorescence micrographs of a test cryo-EM sample containing the calcium sensing dye Fura Red, with a 0.5 nm thick layer of calcium chloride deposited on the sample. After flash melting with a 50 µs laser pulse, the fluorescence intensity in the revitrified area decreases, indicating that mixing and binding of the calcium ions to the dye have occurred. The inset shows a difference image, normalized to the image intensity before flash melting. **b** Optical micrographs of the same area recorded in reflection. **c,d** Without calcium chloride deposited onto the sample, flash melting does not produce the same contrast changes.

In conclusion, we have detailed the design and operation of a vacuum apparatus for the physical vapor deposition onto cryo-EM samples that we believe will be particularly useful for the field of microsecond time-resolved cryo-EM. We have illustrated the operation of the instrument by enclosing cryo-EM samples between vapor deposited silicon dioxide membranes and have determined the minimum membrane thickness required for such a geometry to function as an ultrathin liquid cell in laser flash melting experiment. While we have introduced these liquid cells to extend the temporal observation window of microsecond time-resolved cryo-EM experiments,[15] it is conceivable that the fabrication process could also be adapted to enable high-resolution imaging for traditional *in situ* liquid cell observations.[26–28] Our instrument also promises to open up new avenues for initiating protein dynamics through the deposition of reagents that will mix with the cryo-EM sample once it is flash melted. Here, we have illustrated this concept, showing that if we deposit a soluble calcium salt, it mixes with the sample upon laser melting. This promises to significantly extend the range of phenomena that can be observed in microsecond time-resolved cryo-EM experiments. In the future, this principle could be extended to larger molecules that cannot be evaporated without decomposition, such as peptides or entire proteins, with a soft-landing mass spectrometry approach. The compounds are transferred into the gas phase with electrospray ionization, the molecules of interest are selected with a mass filter, and gently deposited onto the sample at low beam energies to avoid fragmentation upon impact.[13] In broader terms, we envision that our instrument will form the basis for an integrated platform for microsecond time-resolved cryo-EM experiments. Such an instrument will combine the ability to deposit various compounds onto cryo-EM samples with an optical microscope to perform laser flash melting experiments in the same instrument.[7] This will reduce sample contamination by decreasing the number of sample transfers required, and enable complex experiments in which the deposition of different components is alternated with laser revitrification steps.

**Acknowledgments:**

This work was supported by the Duke Center for Structural Biology, NIH grant U54AI170752 and the Swiss National Science Foundation Consolidator Grant TMCG-2_213773. ChatGPT and Perplexity were used for editing the language of the manuscript for clarity, while the scientific content is entirely our own work.

**Author contributions:**

**Wyatt A Curtis:** Data Curation (lead); Formal Analysis (lead); Methodology (equal); Visualization (equal); Writing – original draft (equal); Writing – review & editing (equal). **Constantin R. Krüger:** Data Curation (supporting); Formal Analysis (supporting); Methodology (equal); Visualization(equal); Writing – original draft (equal); Writing – review & editing (equal). **Axel P. Tracol Gavard:** Data Curation (supporting); Methodology (supporting); Writing – review & editing (equal). **Jakub Wenz:** Data Curation (supporting); Methodology (equal); Writing – review & editing (equal). **Marcel Drabbels:** Supervision (supporting) Writing – review & editing (supporting). **Ulrich J. Lorenz:** Conceptualization (lead); Supervision (lead); Funding Acquisition (lead); Project administration (lead); Visualisation (equal); Writing – original draft (equal); Writing – review & editing (equal).

**Data Availability:**

The data that support the findings of this study are available within the article.

**Competing interests:**

The authors have filed for a patent.

- Patent application US 63/767,702 “High-resolution liquid cells for microsecond time-resolved cryo-EM” filed on 12.03.2025

**References**

1. Williams, R. C. & Wyckoff, R. W. G. Applications of Metallic Shadow-Casting to Microscopy. *J. Appl. Phys.* **17**, 23–33 (1946).
2. Hermann, R., Pawley, J., Nagatani, T. & Müller, M. Double-Axis Rotary Shadowing for High-Resolution Scanning Electron Microscopy. *Scanning Microscopy* **2**, (1988).
3. Goldstein, J. I. *et al.* Coating Techniques for SEM and Microanalysis. in *Scanning Electron Microscopy and X-Ray Microanalysis: A Text for Biologist, Materials Scientist, and Geologists* (eds Goldstein, J. I. et al.) 461–494 (Springer US, Boston, MA, 1981). doi:10.1007/978-1-4613-3273-2_10.
4. Wagner, F. R. *et al.* Preparing samples from whole cells using focused-ion-beam milling for cryo-electron tomography. *Nat Protoc* **15**, 2041–2070 (2020).
5. Lorenz, U. J. Microsecond time-resolved cryo-electron microscopy. *Current Opinion in Structural Biology* **87**, 102840 (2024).
6. Voss, J. M., Harder, O. F., Olshin, P. K., Drabbels, M. & Lorenz, U. J. Rapid melting and revitrification as an approach to microsecond time-resolved cryo-electron microscopy. *Chemical Physics Letters* **778**, 138812 (2021).
7. Bongiovanni, G., Harder, O. F., Drabbels, M. & Lorenz, U. J. Microsecond melting and revitrification of cryo samples with a correlative light-electron microscopy approach. *Front. Mol. Biosci.* **9**, (2022).
8. Straub, M. S. *et al.* Laser flash melting cryo-EM samples to overcome preferred orientation. *Nat Methods* 1–7 (2025) doi:10.1038/s41592-025-02796-y.
9. Glaeser, R. M. & Han, B.-G. Opinion: hazards faced by macromolecules when confined to thin aqueous films. *Biophys Rep* **3**, 1–7 (2017).
10. Drulyte, I. *et al.* Approaches to altering particle distributions in cryo-electron microscopy sample preparation. *Acta Cryst D* **74**, 560–571 (2018).
11. Lyumkis, D. Challenges and opportunities in cryo-EM single-particle analysis. *Journal of Biological Chemistry* **294**, 5181–5197 (2019).
12. Hirst, I. J., Thomas, W. J. R., Davies, R. A. & Muench, S. P. CryoEM grid preparation: a closer look at advancements and impact of preparation mode and new approaches. *Biochem Soc Trans* **52**, 1529–1537 (2024).

13. Barrass, S. V. *et al.* Cryo-EM Sample Preparation with Soft-Landing and Laser Flash Melting. 2025.06.05.657968 Preprint at https://doi.org/10.1101/2025.06.05.657968 (2025).

14. Mertz, K. L. *et al.* Laser-Induced Rehydration of Cryo-Landed Proteins Restores Native Structure. *Molecular & Cellular Proteomics* **24**, (2025).

15. Curtis, W. A. *et al.* Ultrathin Liquid Cells for Microsecond Time-Resolved Cryo-EM. 2025.05.05.652279 Preprint at https://doi.org/10.1101/2025.05.05.652279 (2025).

16. Harder, O. F., Barrass, S. V., Drabbels, M. & Lorenz, U. J. Fast viral dynamics revealed by microsecond time-resolved cryo-EM. *Nature Communications* **14**, 5649 (2023).

17. Ellis-Davies, G. C. R. Caged compounds: photorelease technology for control of cellular chemistry and physiology. *Nat Methods* **4**, 619–628 (2007).

18. Kolasinski, K. W. Surface and Adsorbate Structure. in *Surface Science: foundations of catalysis and nanoscience* 9–49 (John Wiley & Sons, Ltd, 2012). doi:https://doi.org/10.1002/9781119941798.ch1.

19. Krüger, C. R., Mowry, N. J., Bongiovanni, G., Drabbels, M. & Lorenz, U. J. Electron diffraction of deeply supercooled water in no man's land. *Nat Commun* **14**, 2812 (2023).

20. Malis, R. F., T. ;. Cheng, S. C. ;. Egerton. EELS Log-Ratio Technique for Specimen-Thickness Measurement in the TEM. *J. Electron Microsc. Tech.* **8**, 193–200 (1988).

21. Yesibolati, M. N. *et al.* Electron inelastic mean free path in water. *Nanoscale* **12**, 20649–20657 (2020).

22. Lee, C., Ikematsu, Y. & Shindo, D. Measurement of mean free paths for inelastic electron scattering of Si and SiO2. *Journal of Electron Microscopy* **51**, 143–148 (2002).

23. Yonekura, K., Braunfeld, M. B., Maki-Yonekura, S. & Agard, D. A. Electron energy filtering significantly improves amplitude contrast of frozen-hydrated protein at 300 kV. *Journal of Structural Biology* **156**, 524–536 (2006).

24. Floto, R. A., Mahaut-Smith, M. P., Somasundaram, B. & Allen, J. M. IgG-induced Ca2+ oscillations in differentiated U937 cells; a study using laser scanning confocal microscopy and co-loaded Fluo-3 and Fura-Red fluorescent probes. *Cell Calcium* **18**, 377–389 (1995).

25. AAT Bioquest, Inc. Spectrum [Fura Red (calcium bound)]. (2026).

26. Ross, F. M. Opportunities and challenges in liquid cell electron microscopy. *Science* **350**, aaa9886 (2015).

27. Pu, S., Gong, C. & Robertson, A. W. Liquid cell transmission electron microscopy and its applications. *Royal Society Open Science* **7**, 191204 (2020).

28. Tarnawski, T. & Parlińska-Wojtan, M. Opportunities and Obstacles in LCTEM Nanoimaging – A Review. *Chemistry–Methods* **4**, e202300041 (2024).